\documentclass[useAMS,usenatbib]{mn2e}
\usepackage{graphicx}

\title[Galactic structure studies from BATC survey]{Galactic structure studies from BATC survey}
\author[Cuihua Du et al.]{Cuihua Du$^{1,2}$,
\thanks{E-mail:ducuihua@gucas.ac.cn}
 Jun Ma$^{2}$, Zhenyu Wu$^{2}$ and Xu Zhou$^{2}$ \\
$^{1}$College of Physical Sciences, Graduate university of the
          Chinese Academy of Sciences, Beijing 100049, P. R. China\\
$^{2}$National Astronomical Observatories, Chinese
              Academy of Sciences, Beijing 100012, P. R. China}

\begin{document}

\date{Received}

\pagerange{\pageref{firstpage}--\pageref{lastpage}} \pubyear{2002}

\maketitle

\label{firstpage}

\begin{abstract}

We present an analysis of the photometric parallaxes of stars in
21 BATC fields carried out with the National Astronomical
Observatories (NAOC) $60/90$ cm Schmidt Telescope in 15
intermediate-band filters from 3000 to 10000 {\AA}. In this study,
we have adopted a three-component (thin disk, thick disk and halo)
model to analyze star counts information. By calculating the
stellar space density as a function of distance from the Galactic
plane, we determine that the range of scale height for the thin
disk varies from 220 to 320 pc. Although 220 pc seems an extreme
value, it is close to the lower limit in the literature. The range
of scale height for the thick disk is from 600 to 1100 pc, and the
corresponding space number density normalization is 7.0-1.0$\%$ of
the thin disk. We find that the scale height of the disk may be
variable with observed direction, which cannot simply be
attributed to statistical errors. Possibly the main reasons can be
attributed to the disk (mainly the thick disk) is flared, with a
scale height increasing with radius. The structure is consistent
with merger origin for the thick disk formation. Adopting a de
Vaucouleurs $r^{1/4}$ law halo, we also find that the axis ratio
towards the Galactic center is somewhat flatter ($\sim 0.4$),
while the shape of the halo in the anticentre and antirotation
direction is rounder with $c/a> 0.4$. Our results show that star
counts in different lines of sight can be used directly to obtain
a rough estimate of the shape of the stellar halo.  Our solutions
support the Galactic models with a flattened inner halo, possibly
it is formed by a merger early in the Galaxy's history.

\end{abstract}

\begin{keywords}
Galaxy: structure-Galaxy: fundamental parameters- Galaxy: stellar
component-Galaxy: halo
\end{keywords}

%___________________________________________________________________
%Section 1
\section{Introduction}

The detailed study of the Galactic structure enables us to address
many important questions in astrophysics, for it is only in the
Milky Way that we can make detailed studies. But, since we observe
our Galaxy from within it, we must use indirect tools such as star
counts to probe its structure (Peiris 2000). The star counts
method, which is predominantly used to study the general
properties of the Galaxy, is a very effective way of constraining
the structural parameters for the components of the Galaxy, while
the density distribution of the Galaxy components are assumed
similar to those of galaxies of the same Hubble type. In the
standard model (Bahcall \& Soneira 1980), the Galaxy is of Hubble
type $Sbc$, consisting of an exponential disk and a spherical
halo.

Over the past decades, considerable efforts have been undertaken
to gain information about the structure and history of formation
and evolution of our Galaxy.  The Galactic structure models of
varying degrees of complexity have been developed (Bahcall  \&
Soneira 1980, 1984; Gilmore 1984; Robin \& Cr\'{e}z\'{e} 1986;
Reid \& Majewski 1993; M\'{e}ndez et al. 1996; Siegel et al.
2002). It is now apparent from several independent avenues of
research that our Milky Way is much more complex system than we
thought before.

Bahcall \& Soneira (1980) established the first standard model, in
which the Galaxy was simplified and parameterized by an
exponential disk and a spheroid, the latter is characterized by a
de Vaucouleurs profile. Later, future studies on this subject
showed that the number of population components of the Galaxy
increased from two to three. This new component (the thick disk)
was introduced by Gilmore \& Reid (1983),  based on star counts
towards the South Galactic pole. The new component is discussed by
Gilmore \& Wyse (1985) and Wyse \& Gilmore (1986). The stellar
population of the thick disk is distinct from that of the halo,
and its existence is seen clearly in colour magnitude diagrams
derived from star count survey (Gilmore et al. 1989, Chen et al.
2001). Following the work of Gilmore \& Reid (1983), three
components model including the thin disk, thick disk and spheroid
(halo) have become a common model for our Galaxy and has been used
widely.

Up to now, the basic stellar components of the Milky Way are the
thin disk, thick disk, stellar halo and central bulge, albeit that
the inter-relationships and distinction amongst different
components remain subject to some debate (Lemon et al. 2004). In
the previous works, various models have been developed to describe
the stellar populations of the Galaxy. In general, these models
were based on the assumption of a suitable spatial density
distribution, and on the observational luminosity function and
colour-magnitude diagram for each stellar population (Bahcall \&
Soneira 1984; Reid \& Majewski 1993) to fit the structural
parameters by exploiting the measurements of colour and
magnitudes. The canonical spatial density distribution is as
follows: stellar distribution for thin disk and thick disk in
cylindrical coordinates by radial and vertical exponentials law
and for the halo by the de Vaucouleurs spheroid (Du et al. 2003;
Karaali et al. 2004). Thus, the structure parameters can then be
deduced by comparing model and star count data. Different
parametrization of the Galactic components  were tried by many
authors (Bahcall \& Soneira 1980; Gilmore 1984; Ojha et al. 1996;
Chen et al. 2001; Karaali et al. 2003, 2004; Du et al. 2003;
Kaempf et al. 2005). However, due to different and conflicting
results from modelling of star counts, the spatial distribution of
the Galactic components remain controversial. There is still some
uncertainty about the exact characteristics of each Galactic
component. But quantifying the properties of the stellar
components of the Galaxy is of wide importance; they are closely
related to stellar quantities such as distance, age, metallicity,
and kinematics characteristics, which are necessary for
understanding the formation and evolution of the Galaxy (Du et al.
2004b; Pohlen et al. 2004; Brook et al. 2005). These properties
can be obtained by straightforward photometric and spectroscopic
observation. At present,  before the full exploitation of the huge
spectral surveys (e.g. GAIA, SEGUE, LAMOST, etc) is possible, star
counts based on all-sky photometric surveys is one of the few
accessible methods for the study of the Galactic structure.

In this study, to better understand and study the Galactic
structure, the Beijing-Arizona-Taiwan-Connecticut (BATC)
multi-colour photometric survey further provides more new
catalogues with the achievement of data observation and reduction.
These catalogues are very useful in constraining the structure of
the main components of the Galaxy. We will report our
investigation of star count extending our research to include
different direction data. The present discussion is a more
complete and sophisticated investigation of a number of BATC
selected fields. Section 2 describes the details of observations
and data reduction.  The object classification and the photometric
parallaxes are found in Sect. 3. Sect. 4 deals with the space
density distribution of stars in the direction used in the study.
Finally, we summarize and discuss our main conclusions.

%___________________________________________________________________
%section 2
\section{BATC Observations}
\subsection{BATC photometric system and data reduction}

The BATC survey performs photometric observations with a large
field multi-colour system. There are 15 intermediate-band filters
in the BATC filter system, which covers an optical wavelength
range from 3000 to 10000 {\AA} (Fan et al. 1996; Zhou et al.
2001). The 60/90 cm f/3 Schmidt Telescope of National Astronomical
Observatories (NAOC) was used, with a Ford Aerospace
2048$\times$2048 CCD camera at its main focus. The field of view
of the CCD is $58^{\prime}$ $\times $ $ 58^{\prime}$ with a pixel
scale of $1\arcsec{\mbox{}\hspace{-0.15cm}.} 7$.

The definition of magnitude for the BATC survey is in the AB
system, which is a monochromatic flux system first introduced by
Oke \& Gunn (1983). The 4 Oke \& Gunn (1983) standards which are
used for flux calibration in the BATC survey are HD19445, HD84937,
BD+262606 and BD+174708. The fluxes of the four stars have been
recalibrated by Fukugita et al. (1996). Their magnitudes in the
BATC system have slightly been  corrected by cross-checking with
data obtained on a number of photometric nights (Zhou et al.
2001).

Preliminary reductions of CCD frames, including bias subtraction
and flat-fielding correction, were carried out with an automatic
data reduction procedure called PIPELINE I, which has been
developed for the BATC survey (Fan et al. 1996). The HST Guide
star catalogue (GSC) (Jenkner et al. 1990) was then used for
coordinate determination.

A PIPELINE II program based on the DAOPHOT II stellar photometric
reduction package of Stetson (1987) was used to measure the
magnitudes of point sources in the BATC CCD frames. The PIPELINE
II reduction procedure was performed on each single CCD frame to
get the point spread function (PSF) magnitude of each point
source. The magnitudes were then calibrated to the BATC standard
system (Zhou et al. 2003). The other sources of photometric error,
including photo star and sky statistics, readout noise, random and
systematic error from bias subtraction and flat fielding, and the
PSF fitting, are all considered in PIPELINE II. The total
estimated errors of each star are given in the final catalogue
(Zhou et al. 2003). Stars that are detected in at least three
filters are included in the final catalogue.

In Table 1,  we list the parameters of the BATC filters. Col. (1)
and col. (2) represent the ID of the BATC filters,
 col. (3) and col. (4) the central wavelengths
and FWHM of the 15 BATC filters, respectively.

% Table 1_______________________________
   \begin{table}
     \begin{minipage}{120mm}
     \caption{Parameters of the BATC filters }
         \begin{tabular}{cccc}\hline
            \hline
           No. & Filter &  Wavelength  & FWHM \\
                &         & (\AA)   & (\AA)   \\

           \hline

1  & $a$ & 3371.5  & 359    \\
2  & $b$ & 3906.9  & 291    \\
3  & $c$ & 4193.5  & 309    \\
4  & $d$ & 4540.0  & 332    \\
5  & $e$ & 4925.0  & 374    \\
6  & $f$ & 5266.8  & 344     \\
7  & $g$ & 5789.9  & 289    \\
8  & $h$ & 6073.9  & 308   \\
9  & $i$ & 6655.9  & 491   \\
10 & $j$ & 7057.4  & 238   \\
11 & $k$ & 7546.3  & 192   \\
12 & $m$ & 8023.2  & 255   \\
13 & $n$ & 8484.3  & 167   \\
14 & $o$ & 9182.2  & 247    \\
15 & $p$ & 9738.5  & 275   \\

 \hline
\hline
\end{tabular}
\end{minipage}
\end{table}

\subsection{The directions }

Most previous investigations have focused upon one or a few
selected lines-of-sight directions generally either in small areas
to great depth or over a large area to shallower depth (e.g.,
Gilmore \& Reid 1983; Bahcall \& Soneira 1984; Reid \& Majewski
1993; Reid et al. 1996). The deep fields are small with
corresponding poor statistical weight, and the large fields are
limited with shallower depth which may not be able to probe the
Galaxy at large distance (Karaali et al. 2004). However,
investigation into Galactic structure such as quantifying the
properties of the stellar components and substructure of the
Galaxy  obviously benefit from large scale surveys. In addition,
evaluation of star counts in a single direction can lead to
degenerate density-law solution. For instance, increasing the
normalization of the Galactic spheroid and decreasing its axis
ratio represent a degeneracy. This means that in most directions,
one cannot distinguish between models with a flattened spheroid
plus a low normalization ratio of spheroid stars to disk stars,
and those with a high axis ratio plus a high normalization (Peiris
2000). Up to now, only a few programs survey the Galaxy in
multiple directions such as the Basle Halo Program (Buser et al.
1999), the Besancon program (Robin et al. 1996, 2003), the
APS-POSS program (Larsen \& Humphreys 1996), and the SDSS (Chen et
al. 2001)

In this paper, the BATC photometry survey presented 21 selected
fields in the multiple directions. Each field of view is $\sim1$
deg$^{2}$. Table 2 lists the locations of the observed fields
toward the Galactic center and their general characteristics. In
Table 2, Column 1 represents the BATC field name; Columns 2 and 3
represent the right ascension and declination, respectively;
Column 4 represents the epoch; Columns 5 and 6 represent the
Galactic longitude and latitude; and the last two columns
represent the mean reddening and limit magnitude, respectively.
The mean reddening [$E(B-V)$] were determined using the maps of
Burstein \& Heiles (1982). It is obvious that the effects of
interstellar extinction  is small for most fields. As shown in the
Table 2, the most photometric depth of our data is 21.0 mag. in
$i$ band.

The fields used in this paper are towards the Galactic center, the
anticentre, the antirotation direction at median and high
latitudes, $|b|>35^{\circ}$. The fields towards the anticentre and
antirotation directions constrain the structure parameters towards
the outer part of the Galaxy, and the fields towards the Galactic
center constrain the inner part of the Galaxy. Since star counts
at high Galactic latitudes are not strongly related to the radial
distribution, they are well suited to study the vertical
distribution of the Galaxy.

 %  Table 2_______________________________
   \begin{table*}
    \centering
     \begin{minipage}{140mm}
     \caption{Direction and relative information for the BATC Galactic structure fields}
         \begin{tabular}{cccccccc}
          \hline
            \hline
Observed field & R.A. & Decl. & epoch& $l$ (deg) & $b$ (deg) & $E(B-V)$ & $i$ (Comp) \\
 \hline

T485 &    8:38:02.00 &    44:58:38.0 &  1950. &    175.7 &    37.8 &   0.03 & 21.0\\
T518 &    9:54:05.60 &    -0:13:24.4 &  1950. &    238.9 &    39.8 &   0.03 & 19.5\\
T288 &    8:42:30.50 &    34:31:54.0 &  1950. &    189.0 &    37.5 &   0.02 & 20.0\\
T477 &    8:45:48.00 &    45:01:17.0 &  1950. &    175.7 &    39.2 &   0.03 & 20.0\\
T328 &    9:10:57.30 &    56:25:49.0 &  1950. &    160.3 &    41.9 &   0.03 & 19.5\\
T349 &    9:13:34.60 &     7:15:00.5 &  1950. &    224.1 &    35.3 &   0.03 & 20.5\\
TA26 &    9:19:57.12 &    33:44:31.3 &  2000. &    191.1 &    44.4 &   0.01 & 20.0\\
T291 &    9:32:00.30 &    50:06:42.0 &  1950. &    167.8 &    46.4 &   0.01 & 20.0\\
T362 &   10:47:55.40 &     4:46:49.0 &  1950. &    245.7 &    53.4 &   0.03 & 20.0\\
T330 &   11:58:02.90 &    46:35:29.0 &  1950. &    147.2 &    68.3 &   0.00 & 20.5\\
U085 &   12:56:04.35 &    56:53:36.6 &  1996. &    121.6 &    60.2 &   0.01 & 21.0\\
T521 &   21:39:19.48 &     0:11:54.5 &  1950. &     56.1 &   -36.8 &   0.06 & 20.5\\
T491 &   22:14:36.10 &    -0:02:07.6 &  1950. &     62.9 &   -44.0 &   0.03 & 20.0\\
T359 &   22:33:51.10 &    13:10:46.0 &  1950. &     79.7 &   -37.8 &   0.06 & 20.5\\
T350 &   11:36:44.16 &    12:14:45.4 &  1950. &    251.3 &    67.3 &   0.00 & 19.5\\
T534 &   15:14:34.80 &    56:30:33.0 &  1950. &     91.6 &    51.1 &   0.01 & 21.0\\
T193 &   21:55:34.00 &     0:46:13.0 &  1950. &     59.8 &   -39.7 &   0.05 & 20.0\\
T516 &    0:52:50.08 &     0:34:52.6 &  1950. &    125.0 &   -62.0 &   0.00 & 20.0\\
T329 &    9:53:13.30 &    47:49:00.0 &  1950. &    169.9 &    50.4 &   0.01 & 21.0\\
TA01 &    0:46:26.60 &    20:29:23.0 &  2000. &    135.7 &   -62.1 &   0.00 & 20.5\\
T517 &    3:51:43.04 &     0:10:01.6 &  1950. &    188.6 &   -38.2 &   0.12 & 20.0\\

\hline
 \end{tabular}
\end{minipage}

{\fontsize{8}{12} NOTES.--- Units of right ascension are hours,
minutes, and seconds, and units of declination are degrees,
arcminutes, and areseconds. }
\end{table*}

%%%%%%%%%%%%%%%%%%%%%%%%%%%%%%%

%Figure 1---------------------------------------

\begin{figure*}
\includegraphics[angle=-90,width=120mm]{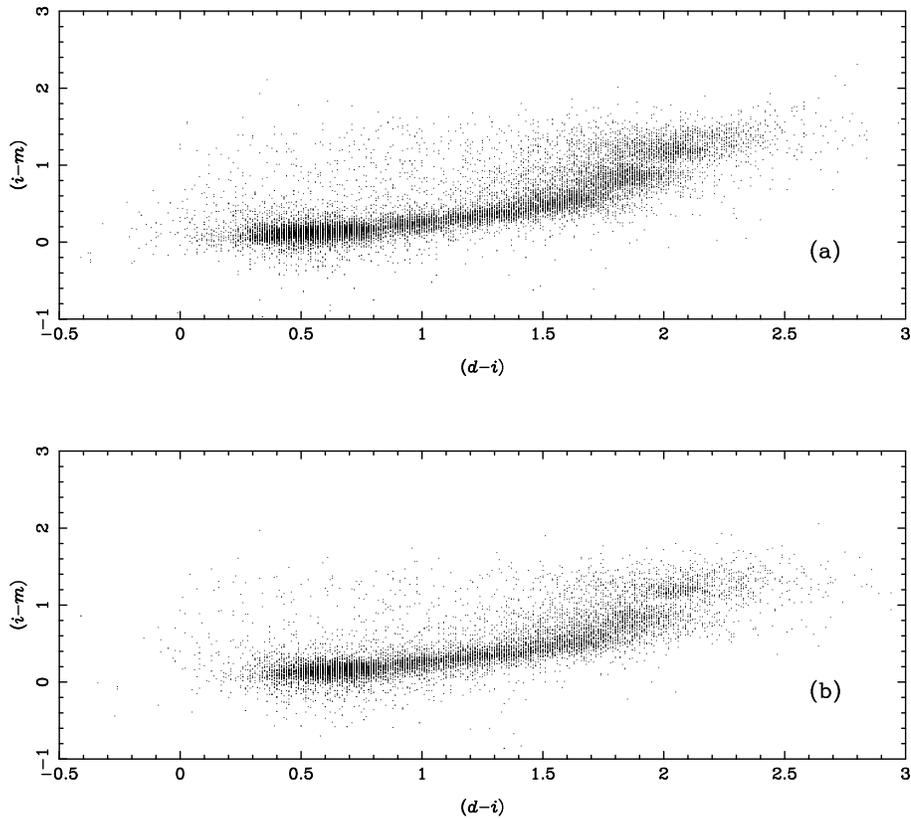}
\caption{The distribution of ($i-m$) versus ($d-i$) for (a) the
northern sample (north of the Galactic plane), (b) southern sample
(south of the Galactic plane) down to the limiting magnitude.}
\end{figure*}

\section{Object classification and photometric parallaxes}
Since there are 15 intermediate-band filters with an optical
wavelength range from 3000 to 10000 {\AA} in the BATC multi-colour
system. Every object observed in all BATC fields could be
classified according to their SED information constructed from the
15-colour photometric catalogue. Here, because our fields in this
work have also been observed by the Sloan Digital Space Survey
(SDSS-DR4) and each object type (stars-galaxies-QSO) has been
given. Thus, we can obtain a relative reliable star catalogue.

The observed colour of each star are compared with a colour
library of known stars with the same photometric system. The input
library for stellar spectra is the Pickles (1998) catalogue. This
library consists of 131 flux-calibrated spectra, including all
normal spectral types and luminosity classes at solar abundance,
and metal-poor and metal-rich F$-$G dwarfs and G$-$K giant
components. Our sample may contain stars spread over a range of
different metallicities. In Fig. 1, the two-colour diagram $(d-i)$
versus $(i-m)$ is shown for our sample. The panel (a) represents
the northern sample (north of the Galactic plane), and the panel
(b) represents the southern sample (south of the Galactic plane).
Although the 15 filters are used in the object classification, the
two-colour diagram based only on the $d$, $i$, $m$ filters as an
example shows that the scatter still exists in our sample. Most of
stars lie in the mean main sequence track, and for those objects
beyond the track, the scatter can be mainly due to the metallicity
effect and a few non-main sequence stars contamination.

For those stars, the probability of belonging to a certain star
class is computed by the SED fitting method. The standard $\chi^2$
minimization, i.e., computing and minimizing the deviations
between photometric SED of a star and the template SEDs obtained
with the same photometric system, is used in the fitting process.
The minimum $\chi^2_{min}$ indicates the best fit to the observed
SED by the set of template spectra:
\begin{equation}
{\chi^{2}=\sum\limits_{l=1}^{N_{filt}=15}\left [\frac{F_{obs,l}-
 F_{temp,l}-b} {\sigma_{l}} \right ]^{2}},
\end{equation}
where ${F_{obs,l}}$, $F_{temp,l}$ and $\sigma_{l}$ are the
observed fluxes, template fluxes and their observed uncertainty in
filter $l$, respectively, and $N_{filt}$ is the total number of
filters in the photometry, while $b$ is the mean magnitude
difference between observed fluxes and template fluxes. Details
about the classification of stars could be found in our previous
papers about star counts (Du et al. 2003, 2004a).

Thus, we can obtain the spectral types and luminosity classes for
stars in the BATC survey. After knowing the stellar type, the
photometric parallaxes can be derived  by estimating absolute
stellar magnitudes. We adopted the absolute magnitude versus
stellar type relation for main-sequence stars from Lang (1992). A
variety of errors affect the determination of stellar distances.
The first source of errors could be from photometric uncertainty;
the second from the misclassification that affects the derivation
of absolute magnitude. In addition, there may be an error by the
contamination of binary stars in our sample. We neglect the effect
of binary contamination on distance derivation. For binaries with
equal mass components, the distance will be assumed closer by a
factor of $\sqrt{2}$ (Ojha et al. 1996). Due to the unknown but
probable mass distribution in binary components, the effect is
certainly less severe (Majewski 1992; Kroupa et al. 1993). Since
most fields are at intermediate and high latitude
($b>35^{\circ}$), the error from the interstellar extinction in
distance calculation can be neglected.

%Figure 2________________________________________________

\begin{figure*}
\includegraphics[angle=-90,width=120mm]{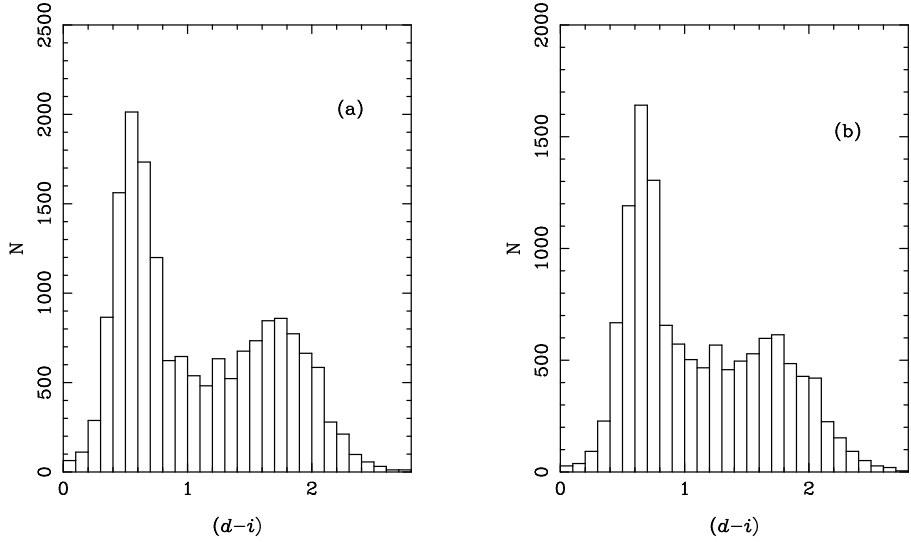}
\caption{The ($d-i$) colour distribution of the sample stars in
the northern and the southern fields.}
\end{figure*}

%Figure 3________________________________________________

\begin{figure*}
   \includegraphics[angle=-90,width=100mm]{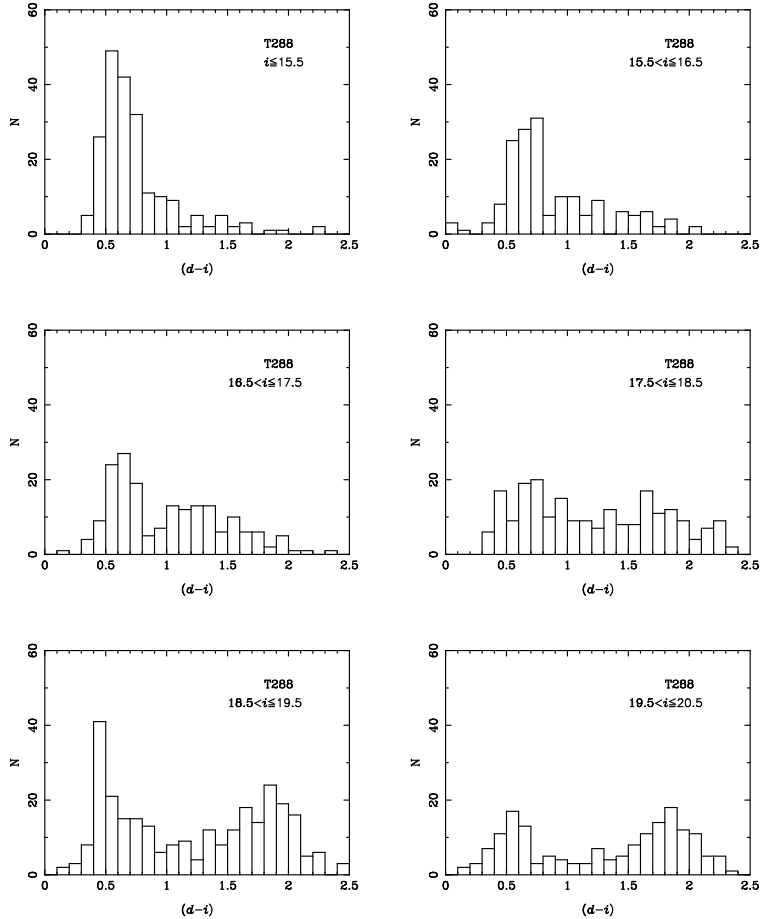}
   \caption{Colour distribution for the T288 field in our sample as a
function of apparent magnitude.}
   \end{figure*}

\section{The stellar density distribution}

It is well known that the population types are a complex function
of both colour and apparent magnitude. Standard star count models
indicate that the colour-magnitude range could be used to separate
roughly different populations of the Galaxy. In Fig. 2, the
($d-i$) colour distribution of the sample stars shows a bimodel
distribution. The left-hand peak is dominated by halo stars, while
the one on the right is dominated by thin disk stars, and the
overlap between the halo stars and the thin disk stars is
dominated by thick disk stars.  Because halo stars are far more
distant than the bulk of the disk stars, only the luminous stars
in the halo (predominantly main-sequence stars) can be detected.
For the main sequence stars, the intrinsically bright ones have
bluer colour. Besides, the halo stars are generally more metal
poor, and hence tend to dominate the blue peak. Similarly, the
thin disk stars would form the red peak and the thick disk stars
would lie between them. From this figure, we can see that the
thick disk and halo populations overlap in the range
($d-i$)$<1.4$.

The colour distribution of the sample stars in the T288 field is
given as a function of apparent magnitude in Fig. 3. According to
Chen et al. (2001), the thick disk has a turnoff of
$(g^{\prime}-r^{\prime})_{0}=0.33$ and it is dominant at bright
apparent magnitudes, $15<g^{\prime}_{0}<18$ mag, whereas the halo
has a turnoff colour at $(g^{\prime}-r^{\prime})_{0}=0.20$ for the
apparent magnitude fainter than $g^{\prime}_{0}\sim18$ mag.
Karaali et al. (2003) also consider that the corresponding turnoff
colour in the $UBVRI$ system are $(B-V)_{0}=0.41$ and 0.53 for
halo and thick disk, respectively. In Fig. 4, we show the observed
colour-magnitude diagram ($i$, $d-i$) from the the northern and
the southern sample. Contrary to the distribution in Fig. 2, the
thick disk and halo stars can be distinguished from the Fig. 3.
For example, the apparent of a halo turnoff is apparent near
$i=17.5$, ($d-i$)=0.7. The turnoffs for the disk and halo in our
sample are fixed (Table 3). Halo stars dominate the absolutely
bright intervals, thick disk stars indicate the intermediate and
the thin disk stars the faint ones.

% Table 3_______________________________
   \begin{table}
   \begin{minipage}{120mm}
     \caption{The colour-magnitude interval for the statistical discrimination of the three components}
         \begin{tabular}{cccc}
            \hline
           stellar populations  & thin disk &  thick disk  & halo \\

              $i$  &         & ($d-i$)   &    \\

           \hline

(13.0-15.5]  &$\ge0.9$  & $<0.9$  &     -  \\
(15.5-16.5]  &$\ge0.9$  & $<0.9$  &     -   \\
(16.5-17.5]  &$\ge1.0$  & $<1.0$  &     -   \\
(17.5-18.5]  & $\ge0.9$ & [0.7-0.9) & $<0.7$    \\
(18.5-19.5]  &$\ge0.9$  &  -         & $<0.9$    \\
(19.5-20.5]  &          & $\ge1.2$    & $<1.2$    \\

\hline

\end{tabular}
\end{minipage}
\end{table}

%Figure 4________________________________________________

\begin{figure*}
\includegraphics[angle=-90,width=100mm]{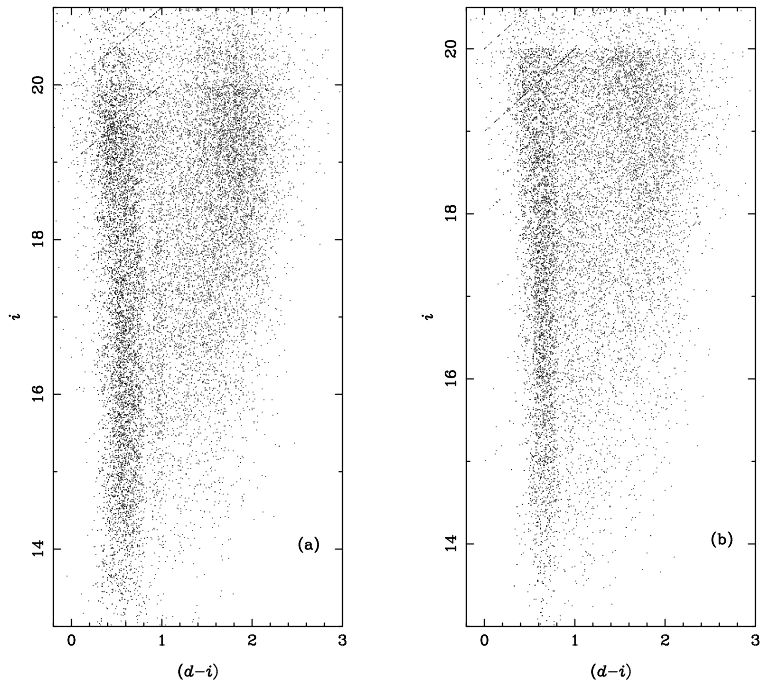}
\caption{Observed ($i$, $d-i$) colour-magnitude diagram from (a)
the northern sample and (b) the southern sample }
\end{figure*}

%Figure 5________________________________________________
\begin{figure}
   \includegraphics[angle=-90,width=90mm]{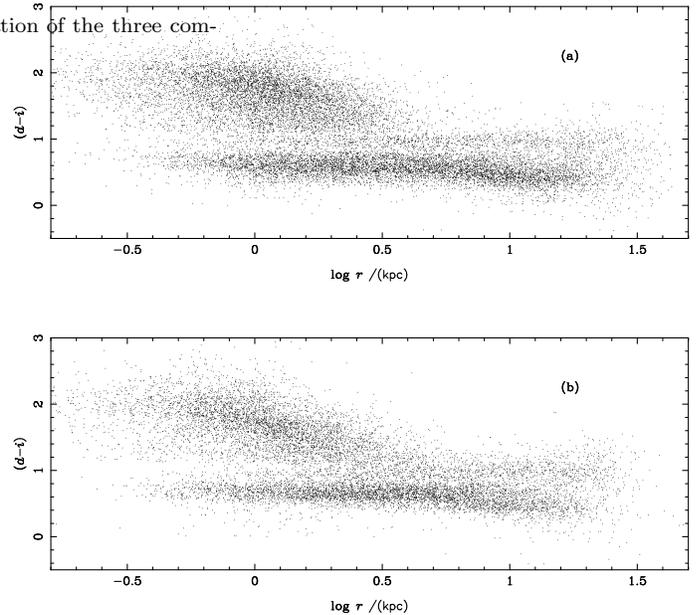}
   \caption{Spatial distribution of ($d-i$), and the photometric distances are derived
  according to the stellar type.}
   \end{figure}

Most of previous studies were based on the assumption of a
suitable spatial density distribution, and on the observational
luminosity function and colour-magnitude diagram for each stellar
population to fit the structural parameters and to interpret them
by simulating the distribution of colour and magnitudes (Gilmore
\& Reid 1983). Here, on account of the use of the photometric
parallaxes, we can make a direct evaluation of the spatial density
law. Rather than trying to fit the structure of the Galaxy in the
observed parameter space of colour and magnitudes, we transfer the
observations into discrete density measurements at various points
in the Galaxy.

In this study, the fields are located at intermediate and high
latitudes, so neither a bulge component nor spiral arms are needed
to describe the observation. The adopted model of stellar density
distribution in this paper includes only two disks (thin disk and
thick disk) and a halo. We try to derive the structural parameters
(e.g. scale height) of the thin and thick disk populations using
our data set. For this we calculate the stellar space density as a
function of distance from the Galactic plane. At first, we use the
two dimensional distribution of stars in the ($d-i$) vs. log$~r$
diagram (Fig. 5) to correct for this incompleteness. The nearest
bins are assumed to be complete; for the incomplete bins we
multiply iteratively by a factor given by the ratio of complete to
incomplete number counts in the previous bin (Phleps et al. 2000).
With the corrected number counts, the density in the log
arithmetic space volume bins $V_j$ can then be calculated
according to
\begin{equation}
\rho_j=\frac{N_j^{corr}}{V_j}
\end{equation}
here, $V_j=(\pi/180)^2(\omega/3)(r_{j+1}^{3}-r_j^3)$ is
partial volume,
$r_{j+1}$ and $r_{j}$ are the limiting distances; and $\omega$ is field size in square degrees.

\subsection{Stellar density distribution in the disk}
We have used a family of standard density laws to describe the
populations of the Milky Way Galaxy. Disk structures are usually
parameterized in cylindrical coordinates by radial and vertical
exponential,
\begin{equation}
\rho(z, r)=\rho_{0}e^{-z/h_{z}}e^{-(x-R_{0})/h_{l}},
 \end{equation}
where $z$ is vertical distance from the Galactic plane, $x$ is the
Galactocentric distance in the plane, $R_{0}$ is is the solar
distance to the Galactic center (8.5 kpc), $\rho_{0}$ is the
normalized local density, $h_{z}$ and $h_{l}$ are the scale height
and scale length of disk, respectively. A similar form uses the
sech$^{2}$ function to parameterize the vertical distribution:
\begin{equation}
\rho(z, r)=\rho_{0}sech^{2}(-z/z_{0})e^{-(z-R_{0})/h_{l}}.
 \end{equation}

A squared secans hyperbolicus is the sum of two exponentials. The
functional form represents a self-gravitating isothermal disk, and
it avoids a singularity at $z=0$. Some studies show the $sech^{2}$
fits better relative to the exponential density for faint absolute
magnitude intervals for the thin disk in the optical star counts
(Gould et al. 1996, Bilir et al. 2005). However, Hammersley et al.
(1999) have shown that an exponential distribution, despite the
singularity problems, is a much better fit to the infrared star
counts of the Galaxy. In addition, Phleps (2000) have shown that
there is a only minor difference between $sech^{2}$ hyperbolicus
and exponential fit when the distance is less than 1 kpc. At large
distances, the $sech^{2}$ function approximates the observed
exponential density profile. We have chosen to use exponential
functional form in this paper. As long as vertical direction is
considered, the equation is as follows,
\begin{equation}
\rho(z)=n_{1}e^{(-z/h_{1})}+n_{2}e^{(-z/h_{2})},
\end{equation}
$h_{1}$ and $h_{2}$ are the scale heights of the thin disk and
thick disk, respectively.

%Figure 6________________________________________________
   \begin{figure*}
   \includegraphics[angle=-90,width=120mm]{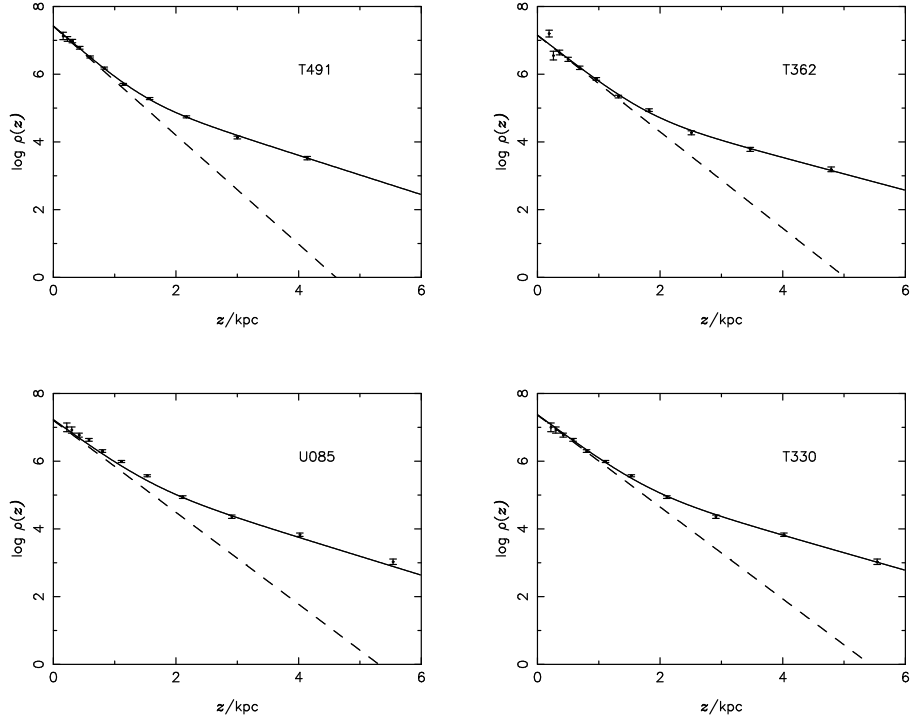}
   \caption{Vertical density distribution of the disk stars in the four field as a example.
The solid line is a fit with a superposition of two exponentials,
the dashed line is the single exponential fit for the thin disk
component.}
   \end{figure*}

Fig. 6 shows the resulting density distribution of the disk stars
in the four fields as an example. The solid line represents a fit
with a superposition of two exponentials, and the dashed line is
the fit for the thin disk component. The comparison between data
and models is made using a $\chi^2$-fit. The most likely values
for thin disk scale height $h_{1}$, thick disk scale height
$h_{2}$ and the corresponding space number density normalization
$n_{2}/n_{1}$ are given in Table 4. Here, the thick disk density
normalization is given in comparison to the density of the thin
disk at the sun. The errors of scale heights and the corresponding
space number density normalization are estimated at a $68\%$
confidential level. We find that the scale height is variable with
the direction. The range of scale height for the thin disk vary
from 220 to 320 pc. An old disk with an exponential scale height
lower than the canonical value (325 pc) has been indicated by some
authors such as Gilmore (1984); Bahcall \& Soneira (1984); Reid \&
Majewski (1993). Although 220 pc seems an extreme value, it is
close to the lower limit in the literature. The range of scale
height for the thick disk is from 600 to 1100 pc, and the
corresponding space number density normalization is $7.0-1.0\%$ of
the thin disk. At the same time, it shows that the thick disk
dominates star counts at distances between 1.5 and 4 kpc over the
galactic plane.

Some results for the thin disk scale height have been published in
the literature. For example, some authors (Gilmore 1984; Bahcall
\& Soneira 1984; Yoshii et al. 1987; Reid \& Majewski 1993)
derived a scale height of 325 pc for old-disk stars; Chen (2001)
derived the scale height of the thin disk to be 330 pc using two
large star count samples. However, Kuijken \& Gilmore (1989)
derived a scale height of 249 pc for the old-disk stars. Haywood
(1994) showed, from an analysis of numerous star counts towards
the pole using his self-consistent evolutionary model, that the
thin disk scale height does not exceed 250 pc. Ojha (1999) also
found that the scale height of the thin disk is 240 pc based on an
analysis of two star count samples. Siegle et al. (2002) gave
apparent scale height Z$_{0,thin}=280-350$ pc. Our result for the
thin disk scale height is in a range of $220-320$ pc. Karaali et
al. (2004) showed that the scale height for thin disk decreases
from absolutely bright to faint stars in a range $265-495$ pc.
They discussed the large range of Galactic structure parameters
and claimed that Galactic model parameters are absolute magnitude
dependent (Bilir et al. 2006).  It is clear that our derived thin
disk scale height is close to the value presented by Siegle et al.
(2002) and Karaali et al.(2004) from several selected areas. This
is not a surprise, since most studies are based on investigation
of one or a few fields in different directions.

%Figure 7________________________________________________
   \begin{figure*}
   \includegraphics[angle=-90,width=120mm]{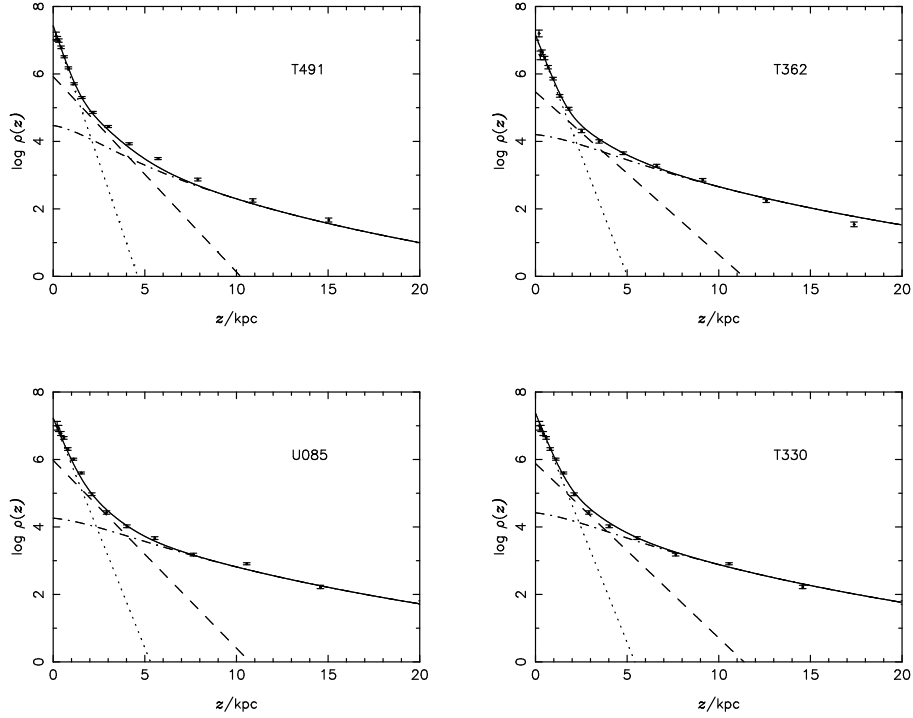}
   \caption{Density distribution perpendicular to the Galactic plane,
the dotted line shows the contribution of the thin disk component,
the dashed line is the contribution of the thick disk,
the dot-dashed line is a de Vaucouleurs law and the solid line the sum of the three.}
   \end{figure*}

The thick disk vertical structure is generally described as
exponential with  scale  heights varying between 480 pc and 1500
pc and its local density is between 1\% and 15\% relative to the
thin disk. Because of the small proportion of the thick disk
locally with regard to the thin disk, it is difficult to derive a
accurate scale height and local density of the thick disk. In
general, any values of $h_{z}$ in the range $480-1500$ pc and of
local density in $1\%- 15\%$ turn out to be acceptable (Robin et
al. 1996). The thick disk's scale height is anticorrelated with
its local density when fitted simultaneously in star count
analysis, and a small scale height is obtained in combination with
high local density, while large scale height is associated with
low local density (Robin et al. 1996).

In some studies, the range of the parameters is large especially
for the thick disk. For example, Gilmore (1984) presented a scale
height of 1300 pc and local normalization of 2\%, Kuijken \&
Gilmore (1989) derived a scale height of 1000 pc and local
normalization of 4\%. Robin et al. (1996) used broad-band
multi-colour photometric and proper motion data to derive a scale
height of $h_{z}=760\pm50$ pc with a local density of
$5.6\pm1.0$\% relative to the thin disk.  Spagna et al. (1996)
used a $BVR$ star count and proper motion data towards the NGP to
derive the scale height of $1137\pm61$pc with a local density of
4.3\%. Ojha et al. (1999) presented a scale height of 790 pc with
a local density of 6.1\% of the thin disk from a photometry and
proper-motion survey in the two directions at intermediate
latitude.  Chen (2001) gave a thick disk scale height between 580
pc and 750 pc, with a local density of $13-6.5\%$ of the thin
disk. Siegel et al. (2002) also investigated these parameters,
matching their Galactic model against deep multi-colour star count
data for seven fields spinning a range of latitude and longitude.
They derived a thick disk scale height of 740 pc with an 8.5\%
normalization to the old disk.  Our derived results in this work
show the thick disk scale height of 600-1100 pc, with a
corresponding density normalization is $7.0-1.0\%$ of the thin
disk.

In summary, the scale height derived from various studies show
large divergence, which cannot simply be attributed to statistical
errors. There could be a number of reasons why the scale height
varies with the observed direction in our study. The first reason
could be from photometric parallaxes uncertainty arising from
either misclassification or metallicity correction. But, unless
the parallax correction is incorrect, this would not produce the
effects we are seeing. The second possibility is that our adopted
models assume the scale height is constant with radius from the
Galactic center. However, maybe the disk (mainly the thick disk)
is flared, with a scale height that increases with radius. This
possibility is also mentioned in Siegel et al. (2002). In
addition, it is clear that star counts when restricted to a small
number of Galactic directions and a small magnitude range do not
give a strong constraint on the scale height.

% Table 4_______________________________
   \begin{table*}
   \centering
   \begin{minipage}{160mm}
     \caption{The Galactic structure parameters derived from fields in this study.}
         \begin{tabular}{c|c|c|c|c}
            \hline
         field name & thin disk  & thick disk &  thick disk  & halo \\
                    &  $h_{1}$ (pc) & $h_{2}$ (pc) &  local normalization & axis ratio \\

           \hline

T485&      $310\pm17$&  $930\pm20$&   0.010&  0.70\\
T518&      $220\pm17$&  $600\pm15$&   0.026&  0.37\\
T288&      $265\pm20$&  $810\pm25$&   0.021&  0.67\\
T477&      $280\pm25$&  $1020\pm25$&  0.010&  0.67\\
T328&      $245\pm25$&  $720\pm15$&   0.030&  0.46\\
T349&      $260\pm16$&  $600\pm30$&   0.026&  0.49\\
TA26&      $230\pm17$&  $690\pm30$&   0.021&  0.50\\
T291&      $290\pm23$&  $810\pm25$&   0.019&  0.58\\
T362&      $305\pm30$&  $900\pm30$&   0.021&  0.61\\
T330&      $320\pm12$&  $840\pm30$&   0.033&  0.58\\
U085&      $320\pm15$&  $780\pm23$&   0.059&  0.64\\
T521&      $305\pm20$&  $810\pm20$&   0.012&  0.43\\
T491&      $270\pm20$&  $750\pm30$&   0.032&  0.40\\
T359&      $255\pm30$&  $600\pm20$&   0.039&  0.40\\
T350&      $310\pm19$&  $1050\pm30$&  0.024&  0.64\\
T534&      $300\pm20$&  $720\pm20$&   0.048&  0.58\\
T193&      $250\pm30$&  $600\pm20$&   0.069&  0.40\\
T516&      $240\pm20$&  $750\pm25$&   0.028&  0.50\\
T329&      $320\pm15$&  $640\pm30$&   0.068&  0.60\\
TA01&      $320\pm20$&  $960\pm15$&   0.019&  0.61\\
T517&      $220\pm10$&  $750\pm30$&   0.010&  0.44\\

\hline
\end{tabular}
\end{minipage}
\end{table*}

%%%%%%%%%%%%%%%%%%%%%%%%%%%%%%%%%%%%%%%%%%%%%%%%%%%%%%%%%%%%%%%%%%%%%%
\subsection{Stellar density distribution in the halo}

Among all of the Galaxy's populations, the halo is traditionally
expected to have changed the least since it formed, and therefore
it provides important clues to the Galaxy's formation and
evolution. The halo is not only less massive than the disk, but
also it occupies a much larger volume than the disk (Larsen \&
Humphreys 2003). According to the studying for the photographic
plates of nearby galaxies, there are numerous forms for the
density law of spheroid components. The de Vaucouleurs law is most
used to describe the surface brightness profile of elliptical
galaxies. The de Vaucouleurs law is an empirical description of
the density distribution of the Galactic halo. The analytic
approximation is:

\begin{eqnarray}
\rho_{s}(R,b,l)&=&\nonumber
\rho_{0}\frac{exp[-10.093(\frac{R}{R_\odot})^{1/4}+10.093]}
{(\frac{R}{R_\odot})^{(7/8)}}  \\
&\nonumber\times&1.25 \frac{exp[-10.093(\frac{R}{R_\odot})^{1/4}
+10.093]}{(\frac{R}{R_\odot})^{(6/8)}}, ~~R<0.03R_{\odot} \\
&\nonumber\times&[1-0.08669/(R/R_\odot)^{1/4}], ~~R\geq0.03R_\odot   \\
\end{eqnarray}
where $R=(x^2+z^2/\kappa^{2})^{1/2}$ is Galactocentric distance,
$\kappa$ is the axis ratio,
$x=(R_{\odot}^2+d^2$cos$^2b-2R_{\odot}d$~cos$b$~cos$l)^{1/2}$,
$z=d$~sin$b$; $R_{\odot}=8$ kpc is the distance of the sun from
the Galactic center, $b$ and $l$ are the Galactic latitude and
longitude; the normalization factor $\rho_{0}$ is usually
expressed as a percentage of the local spatial density of stars.

Other models have used the power-law form,
\begin{eqnarray}
\rho_{s}(R)=\rho_{0}/(a_{0}^{n}+R^{n}),
\end{eqnarray}
where $a_{0}$ is the core radius (an often omitted parameter).

Although Ng et al. (1997) claimed that a halo population described
as a $r^{1/4}$ law predicts less stars than power law at fainter
magnitudes, the choice of halo density is somewhat arbitrary since
the difference between de Vaucouleurs law and power law are subtle
when seen through such a roughly ground lens as star counts. In
our analysis, we use the blue stars (Table 3) to distinguish the
population of halo stars from our sample stars. We adopted a de
Vaucouleurs law for the halo component of the Galaxy and local
density normalization $\rho_{0}=0.125\%$ in the model. Fig. 7
gives the density distribution of all stars in the four fields as
an example. The dotted line shows the contribution of the thin
disk component; the dashed line is the contribution of the thick
disk; the dot-dashed line is a de Vaucouleurs law, and the solid
line is the sum of the three components.  It can also be seen that
the corresponding plots fit the distribution of the halo stars up
to distance of over 15 kpc above the Galactic plane. Our counts
imply that the axis ratio of the stellar halo varies from 0.4 to
0.7.

From our sample fields, T521, T491, T359, T193 and T534 should
belong to the inner part of the halo, while the others lie in the
outer part of the halo according to their longitude and latitude.
The axis ratio versus the longitude distribution is shown in Fig.
8. It is clear that the axis ratio towards the Galactic center is
somewhat flatter ($\sim 0.4$), while the shape of the halo in the
anticentre and antirotation direction is rounder with $c/a> 0.4$.
For T518,  the deviant result may reflect a fluctuation in the
Galactic density distribution, or a systematic error in the
observational data. In summary, star counts in different lines of
sight can be used directly to obtain a rough estimate of the shape
of the stellar halo, although a dependency on models enters
through isolating halo stars. Besides, since our sample fields are
not in the lower latitude areas, which were noted to have
asymmetry (Larsen \& Humphreys 1996; Newberg \& Yanny 2005; Xu et
al. 2006), we cannot detect the triaxial halo distribution.

The apparent discrepancy of the halo axis ratio from various
studies may be due to the multi-component nature of the Galactic
halo (Buser \& Kaeser 1985).  Hartwick (1987) found that the
metal-poor globular clusters and RR Lyrae stars both had a spatial
distribution that was better fitted by two components; the inner
component, which dominates in the solar neighborhood, is flattened
with an axis ratio of $\sim$0.6, while the outer component is
spherical. Kinman et al. (1994) also found evidence for two
components, one significantly flattened which dominates locally,
and one more spherical, in their sample of halo blue horizontal
branch stars. Some studies of the kinematics and abundance of both
field stars and globular clusters show that the halo is better
described as having two sub-populations--a flattened inner halo
and a spherical outer halo (Siegel et al. 2002). Additional
support for dual-halo models can be drawn from the apparent
dichotomy in detailed chemical abundance of halo stars (Nissen \&
Schuster 1997). In a dual-halo model, nearby stars (Larsen \&
Humphreys 1994, 2003; Wyse \& Gilmore 1989; Lemon et al., 2004;
Siegel et al. 2002; this work) are dominated by the flattened
inner halo while distant stars  are dominated by the round outer
halo (Koo et al. 1986; Bahcall \& Soneira 1984; Preston et al.
1991).

The principal contribution of star counts in constraining Galactic
formation scenarios lies in revealing the underlying shape,
chemistry and ages of the stellar population through sophisticated
modelling. Our study of stellar halo provides some support for the
hybrid formation model. The existence of two components may be
evidence that the stellar halo formed by hybrid collapse process
(Wyse 1995; Van den Bergh 1993; Zinn 1993; Norris et al. 1994;
Chiba \& Beers 2000), or that the stellar halo is locally
flattened in response to the disk potential (Binney \& May 1986),
or perhaps reflects the different orbital parameters and internal
structure of disrupted satellite galaxies that were accreted by
the Galaxy to form the stellar halo (Freeman 1987).

%Figure 8________________________________________________
   \begin{figure}
   \includegraphics[angle=-90,width=80mm]{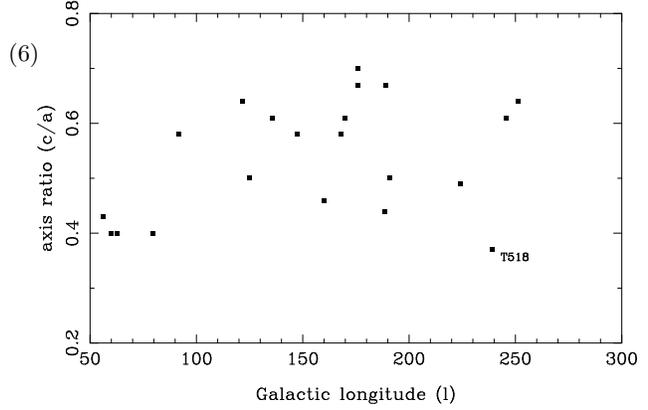}
   \caption{The halo axis ratio distribution versus the the Galactic longitude}
   \end{figure}

\section{Summary and Discussion}

We have analyzed the BATC survey data observed in 21 fields with
the help of a Galaxy model in order to parameterize the vertical
distribution of stars in the Milky Way. The adopted model of the
Galaxy consists of three components: thin disk, thick disk
(exponential form) and halo (de Vaucouleurs law). From the
$\chi^2$ fit to the direct measurement of the stellar density
distribution, we determine that the range of scale height for the
thin disk varies from 220 to 320 pc. Although 220 pc seems an
extreme value, it is close to the lower limit in the literatures.
The range of scale height for the thick disk is from 600 to 1100
pc, and the corresponding space number density normalization is
$7.0-1.0\%$ of the thin disk. Our results show that the scale
height is variable with the observation direction, which cannot be
attributed to statistical errors. Possibly the main reasons can be
attributed to the disk (mainly the thick disk)is flared, with a
scale height that increases with radius. It is consistent with
merger origin for the thick disk formation. The actual numerical
values of Galactic structure parameters are less scientifically
important that what they tell us about the Galaxy in general $-$
for example, the origin of the populations. A number of scenarios
have been proposed for the origin of the thick disk (see review in
Majewski 1993 and Siegel et al. 2002).

In addition, adopting a de Vaucouleurs $r^{1/4}$ law halo and a
local density normalization $\rho_{0}=0.125\%$, we find that the
axis ratio towards the Galactic center is more flatter ($\sim
0.4$), while the shape of the halo in the anticentre and
antirotation direction is rounder with $c/a> 0.4$. It reflects the
shape of the inner halo. In a word, the star counts in different
lines of sight can be used directly to obtain a rough estimate of
the shape of the stellar halo.  With completeness limits for our
selected fields typically 19th to 21th magnitude, our star counts
are most applicable to the inner halo. Our solutions support the
Galactic models with a flattened inner halo. The inner halo is
difficult to distinguish from the thick disk, and it is chemically
and kinematically overlapped the thick disk $-$ possibly formed by
a merger early in the Galaxy's history. The outer halo is more or
less spherical and is disjoint from the inner halo. In particular,
the outer halo might be dominated by substructures that are likely
the remnants of interactions.

Evidence has been growing for some time that simple description of
the Galactic halo are inadequate. It is possible that the axis
ratio may vary with distance, and the halo becomes more spherical
in the outer parts. Some surveys have also found a single axis
ratio too restrictive and adopted a axis ratio ($c/a$) that
increases with Galactocentric radius to explain the stellar
spatial distribution in the halo. The dual halo models may resolve
many of the disagreements in star count results. However,
answering the question of whether or not the multi-component or
triaxial halo is supported requires more data such as kinematics
and chemical abundances analysis. The question will be resolved
with the coming on-line of future projects aimed at spectroscopic
sky surveys such as SEGUE, LAMOST, GAIA, and further photometric
surveys of the southern sky.

\section*{Acknowledgments}

The BATC Survey is supported by the Chinese Academy of Sciences,
the Chinese National Natural Science Foundation under the contract
No. 10473012 and No. 10573020, the Chinese State Committee of
Sciences and Technology. This work was also supported by the GUCAS
president fund yzjj200501.

\label{lastpage}

\end{document}